\pdfoutput=1
\documentclass[sigconf]{acmart}

\usepackage{multirow}

\usepackage{balance}

\def\BibTeX{{\rm B\kern-.05em{\sc i\kern-.025em b}\kern-.08emT\kern-.1667em\lower.7ex\hbox{E}\kern-.125emX}}
    
\copyrightyear{2020}
\acmYear{2020}
\setcopyright{rightsretained}
\acmConference[AIES '20]{Proceedings of the 2020 AAAI/ACM Conference on AI, Ethics, and Society}{February 7--8, 2020}{New York, NY, USA}
\acmDOI{10.1145/3375627.3375857}
\acmISBN{978-1-4503-7110-0/20/02}

\begin{document}
\fancyhead{}

\title[1]{{Should Artificial Intelligence Governance be Centralised?} Design Lessons from History}

\author{Peter Cihon}
\authornote{Equal contribution, order selected at random.}

\affiliation{%
  \institution{Centre for the Governance of AI, Future of Humanity Institute, University of Oxford}
}
\email{petercihon@gmail.com}

\author{Matthijs M. Maas}
\authornotemark[1]
\affiliation{%
  \institution{CILCC, Faculty of Law, \\University of Copenhagen \& \\
  Centre for the Governance of AI, Future of Humanity Institute, University of Oxford}
  }
\email{matthijs.maas@jur.ku.dk}

\author{Luke Kemp}
\authornotemark[1]
\affiliation{%
  \institution{Centre for the Study of Existential Risk, University of Cambridge}
}
\email{ltk27@cam.ac.uk}

%

%
\begin{abstract}
Can effective international governance for artificial intelligence remain fragmented, or is there a need for a centralised international organisation for AI? We draw on the history of other international regimes to identify advantages and disadvantages in centralising AI governance. Some considerations, such as efficiency and political power, speak in favour of centralisation. Conversely, the risk of creating a slow and brittle institution speaks against it, as does the difficulty in securing participation while creating stringent rules. Other considerations depend on the specific design of a centralised institution. A well-designed body may be able to deter forum shopping and ensure policy coordination. However, forum shopping can be beneficial and a fragmented landscape of institutions can be self-organising. Centralisation entails trade-offs and the details matter. We conclude with two core recommendations. First, the outcome will depend on the exact design of a central institution. A well-designed centralised regime covering a set of coherent issues could be beneficial. But locking-in an inadequate structure may pose a fate worse than fragmentation. Second, for now fragmentation will likely persist. This should be closely monitored to see if it is self-organising or simply inadequate.
\end{abstract}

%

%
\maketitle

\section{Introduction}In 2018, Canada and France proposed the International Panel on Artificial Intelligence (IPAI). After being rejected at the G7 in 2019, negotiations shifted to the OECD and are presently ongoing. As the field of AI continues to mature and spark public interest and legislative concern \cite{perrault_ai_2019}, the priority of such governance initiatives reflects the growing appreciation that AI has the potential to dramatically change the world for both good and ill \cite{dafoe_ai_2018}. Research into AI governance needs to keep pace with policy-making and technological change. Choices made today may have long-lasting impacts on policymakers' ability to address numerous AI policy problems \cite{cave_bridging_2019}. Effective governance can promote safety, accountability, and responsible behaviour in the research, development, and deployment of AI systems.

AI governance research to date has predominantly focused at the national and sub-national levels \cite{scherer_regulating_2016,calo_artificial_2017,gasser_layered_2017}. Research into AI {\it global} governance remains relatively nascent (though see \cite{UNIRI2019}). \citeauthor{kemp_high-level_2019} \cite{kemp_high-level_2019} have called for specialised, centralised intergovernmental agencies to coordinate policy responses globally, and others have called for a centralised `International Artificial Intelligence Organisation' \cite{erdelyi_regulating_2018}. Others favour more decentralised arrangements based around `Governance Coordinating Committees', global standards, or existing international law instruments \cite{wallach_agile_2018,cihon_standards_2019,kunz_artificial_2020}.

No one has taken a step back to inquire: what would the history of multilateralism suggest, given the state and trajectory of AI? Should AI governance be centralised or decentralised? `Centralisation', in this case, refers to the degree to which the coordination, oversight and/or regulation of a set of AI policy issues or technologies are housed under a single (global) institution. This is not a binary choice; it exists across a spectrum. Trade is highly (but not entirely) centralised under the umbrella of the WTO. In contrast, environmental multilateralism is much more decentralised.

In this paper, we seek to help the community of researchers, policymakers, and other stakeholders in AI governance understand the advantages and disadvantages of centralisation. This may help set terms and catalyse a much-needed debate to inform governance design decisions. We first outline the international governance challenges of AI, and review early proposed global responses. We then draw on existing literatures on regime fragmentation \cite{biermann_fragmentation_2009} and `regime complexes' \cite{orsini_regime_2013} to assess considerations in centralising the international governance of AI. We draw on the history of other international regimes\footnote{A regime is a set of `implicit or explicit principles, norms, rules and decision-making procedures around which actors' expectations converge in a given area of international relations'\cite[p.186]{krasner_structural_1982}.} to identify considerations that speak in favour or against designing a centralised regime complex for AI. We conclude with two recommendations. First, many trade-offs are contingent on how well-designed a central body would be. An adaptable, powerful institution with a manageable mandate would be beneficial, but a poorly designed body could prove a fate worse than fragmentation. Second, for now there should be structured monitoring of existing efforts to see whether they are they are self-organising or insufficient.

\section{The State of AI Governance}
There is debate as to whether AI is a single policy area or a diverse series of issues. Some claim that AI cannot be cohesively regulated as it is a collection of disparate technologies, with different risk profiles across different applications and industries \cite{stone_artificial_2016}. This is an important but not entirely convincing objection. The technical field has no settled definition for `AI',\footnote{We define `AI' as any machine system capable of functioning `appropriately and with foresight in its environment' \cite[p.13]{nilsson_quest_2009}; see too \cite[p.5]{dafoe_ai_2018}.} so it should be no surprise that defining a manageable scope for AI governance will be difficult. Yet this challenge is not unique to AI: definitional issues abound in areas such as environment and energy, but have not figured prominently in debates over centralisation. Indeed, energy and environment ministries are common at the domestic level, despite problems in setting the boundaries of natural systems and resources.

We contend that there are numerous ways in which a centralised body could be designed for AI governance. For example, a centralised approach could carve out a subset of interlinked AI issues to cover. This could involve focusing on the potentially high-risk applications of AI systems, such as AI-enabled cyberwarfare, lethal autonomous weapons (LAWS), other advanced military applications, or high-level machine intelligence (HLMI).\footnote{`High-level machine intelligence' has been defined as `unaided machines [that] can accomplish every task better and more cheaply than human workers' \cite[p.1]{grace2018will}.} Another approach could govern underlying hardware resources (e.g. large-scale compute resources) or software libraries. We are agnostic on the specifics of how centralisation could or should be implemented, and instead focus on the costs and benefits of centralisation in the abstract. The exact advantages and disadvantages of centralisation are likely to vary depending on the institutional design. This is an important area of further study, particularly once more specific proposals are put forward. However, such work must be grounded in a higher-level investigation of trade-offs in centralising AI governance. It is this foundational analysis which we seek to offer.

Numerous AI issues could benefit from international cooperation. These include the potentially catastrophic applications mentioned above. It also encompasses more quotidian uses, such as AI-enabled cybercrime; human health applications; safety and regulation of autonomous vehicles and drones; surveillance, privacy and data-use; and labour automation. Multilateral coordination could also use AI to tackle other global problems such as climate change \cite{rolnick_tackling_2019}, or help meet the Sustainable Development Goals \cite{vinuesa_role_2019}. This is an illustrative but not exhaustive list of international AI policy issues. 

Global regulation across these issues is currently nascent, fragmented, yet evolving. A wide range of UN institutions have begun to undertake some activities on AI \cite{itu_united_2019}. The bodies covering AI policy issues range across existing organisations including the International Labour Organisation (ILO), International Telecommunication Union (ITU), and UNESCO. This is complemented by budding regulations and working groups across the International Organisation for Standardisation (ISO), International Maritime Organisation (IMO), International Civil Aviation Organisation (ICAO), and other bodies, as well as treaty amendments, such as the updating of the Vienna Convention on Road Traffic to encompass autonomous vehicles \cite{kunz_artificial_2020}, or the ongoing negotiations at the Convention on Certain Conventional Weapons (CCW) on LAWS. The UN System Chief Executives Board (CEB) for Coordination through the High-Level Committee on Programmes has been empowered to draft a system-wide AI capacity building strategy. The High-level Panel on Digital Cooperation has also sought to gather together common principles and ideas for AI relevant areas \cite{high2019panel}. Whether these initiatives bear fruit, however, remains questionable, as many of the involved international organisations have fragmented membership, were not originally created to address AI issues and lack effective enforcement or compliance mechanisms \cite[p.2]{morin_how_2019}.

The trajectory of these initiatives matters. How governance is initially organised can be central to its success. Debates over centralisation and fragmentation are long-lasting and prominent with good reason. How we structure international cooperation can be critical to its success, and most other debates often implicitly hinge on structural debates. Fragmentation and centralisation exist across a spectrum. In a world lacking a global government, some fragmentation will always prevail. But the degree to which it prevails is crucial. We define `fragmentation' as a patchwork of international organisations and institutions which focus on a particular issue area, but differ in scope, membership and often rules \cite[p.16]{biermann_fragmentation_2009}. We define centralisation as an arrangement in which governance of a particular issue lies under the authority of a single umbrella body. A regime complex is a network of three or more international regimes on a common issue area. These should have overlapping membership and cause potentially problematic interactions \cite[p.29]{orsini_regime_2013}. These definitions and terms are by nature normatively loaded. For example, some may find `decentralisation' to be a positive framing, while others may see `fragmentation' to possess negative connotations. Recognising this, we seek to use these terms in a primarily analytical manner. We will use findings from each of these theoretical areas to inform our discussion of the history of multilateral fragmentation and its implications for AI governance. 

\section{Centralisation Criteria: History of Governance Trade-Offs}
In the following discussion, we explore a series of considerations for AI governance. Political power and efficient participation support centralisation. The breadth vs. depth dilemma, as well as slowness and brittleness support decentralisation. Policy coordination and forum shopping considerations can cut both ways.  

\subsection{Political Power}

Regimes embody power in their authority over rules, norms, and knowledge beyond states' exclusive control. A more centralised regime will see this power concentrated among fewer institutions. A centralised, powerful architecture is likely to be more influential against competing international organisations and with constituent states \cite[pp.36-7]{orsini_regime_2013}.

An absence of centralised authority to manage regime complexes has presented challenges in the past. Across the proliferation of Multilateral Environmental Agreements (MEAs) there is no requirement to cede responsibility to the UN Environmental Programme in the case of overlap or competition. This has led to turf wars, inefficiencies and even contradictory policies \cite{biermann_fragmentation_2009}. One of the most notable examples is that of hydrofluorocarbons (HFCs). HFCs are potent greenhouse gases, and yet their use has been encouraged by the Montreal Protocol since 1987 as a replacement for ozone-depleting substances. This has only recently been resolved via the 2015 Kigali Amendment to the Montreal Protocol, which itself has a prolonged implementation period. Similarly, the internet governance regime complex is diffuse. Multiple venues and norms govern technical standards, cyber crime, human rights, and warfare \cite{nye_regime_2014}. Although the UN Internet Governance Forum (IGF) discusses several cross-cutting issues, it does not have a mandate to consolidate even principles, let alone negotiate new formal agreements \cite{mueller_internet_2007}. 

In contrast, other centralised regimes have supported effective management. For example, under the umbrella of the WTO, norms such as the most-favoured-nation principle (equally treating all WTO member states) principle have become the bedrock of international trade. The power and track-record of the WTO is so formidable that it has created a {\it chilling} effect: the fear of colliding with WTO norms and rules has led environmental treaties to self-censor and actively avoid discussing or deploying trade-related measures \cite{eckersley_big_2004}. Both the chilling effect and the remarkably powerful application of common trade rules were not a marker of international trade until the establishment of the WTO. The power of these centralised body has stretched beyond influencing states in the domain of trade, to moulding related issues. 

Political power offers further benefits in governing emerging technologies that are inherently uncertain in both substance and policy impact. Uncertainty in technology and preferences has been associated with some increased centralisation in regimes \cite{koremenos_rational_2001-1}. There may also be benefits to housing a foresight capacity within the regime complex, to allow for accelerated or even proactive efforts \cite{pauwels_new_2019}. Centralised AI governance would enable an empowered organisation to more effectively use foresight analyses to inform policy responses across the regime complex.

\subsection{Supporting Efficiency \& Participation}

Decentralised AI governance may undermine efficiency and inhibit participation. States often create centralised regimes to reduce costs, for instance by eliminating duplicate efforts, yielding economies of scale within secretariats, and simplifying participation \cite{esty_revitalizing_2002}. Conversely, fragmented regimes may force states to spread resources and funding over many distinct institutions, particularly limiting the ability of less well-resourced states or parties to participate fully \cite[p.2]{morin_how_2019}. 

Historically, decentralised regimes have presented cost and related participation concerns. Hundreds of related and sometimes overlapping international environmental agreements can create `treaty congestion'  \cite{anton_treaty_2012}. This complicates participation and implementation for both developed and developing nations \cite{esty_revitalizing_2002}. This includes costs associated with travel to different forums, monitoring and reporting for a range of different bodies, and duplication of effort by different secretariats (ibid.). 

Similar challenges are already being witnessed in AI governance. Simultaneous and globally distributed meetings pose burdensome participation costs for civil society. Fragmented organisations must duplicatively invest in high-demand machine learning subject matter experts to inform their activities. Centralisation would support institutional efficiency and participation.

\subsection{Slowness \& Brittleness of Centralised Regimes}
One potential problem of centralisation lies in the relatively slow process of establishing centralised institutions, which may often be outpaced by the rate of technological change. Another challenge lies in centralised institutions' brittleness after they are established, i.e., their vulnerability to regulatory capture, or failure to react to changes in the problem landscape. 

Establishing new international institutions is often a slow process. For example, the Kyoto Protocol took three years of negotiations to create and then another eight to enter into force. This becomes even more onerous with higher participation and stakes. Under the GATT, negotiations for a 26\% cut in tariffs between 19 countries took 8 months in 1947. The Uruguay round, beginning in 1986, took 91 months to achieve a tariff reduction of 38\% between 125 parties \cite{martin2007so}. International law has been quick to respond to technological changes in some cases, and delayed in others \cite[p.184]{picker_view_2001}. Decentralised efforts may prove quicker to respond to complex, `transversal' issues, if they rely more on informal institutions with a smaller but like-minded membership \cite[pp.2-3]{morin_how_2019}. Centralised AI governance may be particularly vulnerable to sparking lengthy negotiations, because progress on centralised regimes for new technologies tends to be hard if a few states hold clearly unequal stakes in the technology, or if there are significant differences in information and expertise among states or between states and private industry \cite[pp.187-94]{picker_view_2001}. Both these conditions closely match the context of AI technology. Moreover, because AI technology develops rapidly, such slow implementation of rules and principles could lead to certain actors taking advantage by setting {\it de facto} arrangements or extant state practice. 

Even after its creation, a centralised regime can be {\it brittle}; the very qualities that provide it with political power may exacerbate the adverse effects of regulatory capture; the features that ensure institutional stability, may also mean that the institution cannot adapt quickly to unanticipated outside stressors outside its established mission. The regime might break before it bends. The first potential risk is regulatory capture. Given the high profile of AI issue areas, political independence is paramount. However, as illustrated by numerous cases, including undue corporate influence in the WHO during the 2009 H1N1 pandemic \cite{deshman_horizontal_2011}, no institution is fully immune to regime capture, and centralisation may reduce the costs of lobbying, making capture easier by providing a single locus of influence. On the other hand, a regime complex comprising many parallel institutions could find itself vulnerable to capture by powerful actors, who are better positioned than smaller parties to send representatives to every forum. 

Moreover, centralised regimes entail higher stakes. Many issues are in a single basket and thus failure is more likely to be severe if it does occur. International institutions can be notoriously path-dependent and thus fail to adjust to changing circumstances, as seen with the ILO's considerable difficulties in reforming its participation and rulemaking processes in the 1990s \cite{baccaro_pathology_2012}. The public failure of a flagship global AI institution or governance effort could have lasting political repercussions. It could strangle subsequent, more well-conceived proposals in the crib, by undermining confidence in multilateral governance generally or capable governance on AI issues specifically. By contrast, for a decentralized regime complex to similarly fail, all of its component institutions would need to simultaneously `break' or fail to innovate at once.\footnote{We thank Nicolas Mo{\"e}s for this observation.} A centralised institution that does not outright collapse, but which remains ineffective, may become a blockade against better efforts. 

Ultimately, brittleness is not an inherent weakness of centralisation--and indeed depends far more on institutional design details. There may be strategies to `innovation-proof'\cite{maas_innovation-proof_2019} governance regimes. Periodic renegotiation, modular expansion, `principles based regulation', or sunset clauses can also support ongoing reform \cite[pp.29-30]{marchant2011growing}. Such approaches have often proved successful historically, due partially to decentralisation but, importantly, also to particular designs.

\subsection{The Breadth vs. Depth Dilemma}

Pursuing centralisation may create an overly high threshold that limits participation. All multilateral agreements face a trade-off between having higher participation (`breadth') or stricter rules and greater ambition of commitments (`depth'). The dilemma is particularly evident for centralised institutions that are intended to be powerful and require strong commitments from states. 

However, the opposite dynamics of sacrificing depth for breadth can also pose risks. The 2015 Paris Agreement on Climate Change was significantly watered down to allow for the legal participation of the US. Anticipated difficulties in ratification through the Senate led to negotiators opting for a `pledge and review' structure with few legal obligations. Thus, the US could join simply through the approval of the executive \cite{kemp_us-proofing_2017}. In this case, inclusion of the US (which at any rate proved temporary) came at the cost of significant cutbacks on the demands which the regime sought to make of all parties. 

In contrast, decentralisation could allow for major powers to engage in relevant regulatory efforts where they would be deterred from signing up to a more comprehensive package. This has precedence in the history of climate governance. Some claim that the US-led Asia-Pacific Partnership on Clean Development and Climate helped, rather than hindered climate governance, as it bypassed UNFCCC deadlock and secured non-binding commitments from actors not bound by the Kyoto Protocol \cite[pp.259-60]{zelli_fragmentation_2011}.

This matters, as buy-in may prove a thorny issue for AI governance. The actors who lead in AI development include powerful states that are potentially most adverse to global regulation in this area. They have thus far proved recalcitrant in the global governance of security issues such as anti-personnel mines or cyberwarfare. In response, some have already recommended a critical-mass governance approach to the military uses of AI. Rather than seeking a comprehensive agreement, devolving and spinning off certain components into separate treaties (e.g. for LAWS testing standards; liability and responsibility; and limits to operational usage) could instead allow for the powerful to ratify and move forward at least a few of those options \cite{weaver_autonomous_2014}.

The breadth vs. depth dilemma is a trade-off in multilateralism generally. However, it is a particularly pertinent challenge for centralisation. The key benefit of a centralised body would be to be a powerful anchor that ensures policy coordination and coherence, without suffering fragmentation in membership. This dilemma suggests it is unlikely to have both. It will likely need to restrict membership to have teeth, or lose its teeth to have wide participation. A critical mass approach may be able to deliver the best of both worlds. Nonetheless these dilemma poses a difficult knot for centralisation to unravel.

\subsection{Forum Shopping}

Forum shopping may help or hinder AI governance, depending on the particular circumstances. Fragmentation enables actors to choose where and how to engage. Such `forum shopping' may take one of several forms: moving venues, abandoning one organisation, creating new venues, and working across multiple organisations to sew competition between them \cite{braithwaite_global_2000}. Even when there is a natural venue for an issue, actors have reasons to forum-shop. For instance, states may look to maximise their influence, appease domestic pressure \cite{pekkanen_trading_2007} and placate constituents by shifting to a toothless forum \cite{helfer_regime_2004}.

The ability to successfully forum-shop depends on an actor's power. Most successful examples of forum-shifting have been led by the US \cite{braithwaite_global_2000}. Intellectual property rights in trade, for example, was subject to prolonged, contentious forum shopping. Developed states resisted attempts of the UN Conference on Trade and Development (UNCTAD) to address intellectual property rights in trade by trying to push them onto the World Intellectual Property Organization (WIPO) (ibid., 566) and then subsequently to the WTO \cite{helfer_regime_2004}, overruling protests from developing states. Outcomes often reflect power, but weak states and non-state actors can also pursue forum shopping strategies in order to challenge the status-quo \cite{jupille_institutional_2013}.

Forum shopping may help or hurt governance. This is evident in current efforts to regulate LAWS. While the Group of Governmental Experts has made some progress, on the whole the CCW has taken slow deliberations on LAWS. In response, frustrated activists have threatened to shift to another forum, as happened with the Ottawa Treaty that banned landmines \cite{delcker_how_2019}. This strategy could catalyse progress, but also brings risks of further forum shopping and weak or unimplemented agreements. Forum shopping may similarly delay, stall, or weaken regulation of time-sensitive AI policy issues, including potential future HLMI development. It is plausible that leading AI firms also have sway when they elect to participate in some venues but not others. The OECD Expert Group on AI included representatives from leading firms, whereas engagement at UN efforts, including the Internet Governance Forum (IGF), do not appear to be similarly prioritised. A decentralised regime will enable forum shopping, though further work is needed to determine whether this will help or hurt governance outcomes on the whole.

\subsection{Policy Coordination}

There are good reasons to believe that either centralisation or fragmentation could enhance coordination. A centralised regime can enable easier coordination both across and within policy issues, acting as a focal point for states. Others argue that this is not always the case, and that fragmentation can mutually supportive and even more creative institutions.

Centralisation reduces the occurrence of conflicting mandates and enables communication. These are the ingredients for policy coherence. As noted previously, the WTO has been remarkably successful in ensuring coherent policy and principles across the realm of trade, and even into other areas such as the environment. 

However, fragmented regimes can often act as complex adaptive systems. Political requests and communication between secretariats often ensures  bottom-up coordination even in the absence of centralisation. Multiple organisations have sought to reduce greenhouse gas emissions within their respective remits, often at the behest of the UNFCCC Conference of Parties. When effective, bottom-up coordination can slowly evolve into centralisation. Indeed, this was the case for the GATT and numerous regional, bilateral and sectoral trade treaties, which all coalesced together into the WTO. While this organic self-organisation has occurred, it has taken decades, with forum shopping and inaction prevailing for many years. 

Indeed, some have argued that decentralisation does not just deliver `good enough' global governance \cite{patrick_unruled_2014} that reflects a demand for diverse principles in a multipolar world. Instead, they argue `polycentric'  governance approaches \cite{ostrom_polycentric_2010} may be more creative and legitimate than centrally coordinated regimes. Arguments in favour of polycentricity include the notion that it enables governance initiatives to begin having impacts at diverse scales, and that it enables experimentation with diverse policies and approaches, learning from experience and best practices (ibid., 552). Consequently, these scholars assume “that the invisible hand of a market of institutions leads to a better distribution of functions and effects” \cite[p.7]{zelli_introduction:_2013}.

It is unclear if the different bodies covering AI issues will self-organise or collide. Many of the issues are interdependent and will need to be addressed in tandem. Some particular policy-levers, such as regulating computing power or data, will impact almost all use areas, given that AI progress and use is closely tied to such inputs. Numerous initiatives on AI and robotics are displaying loose coordination \cite{kunz_artificial_2020}, but it remains uncertain whether the virtues of a free market of governance will prevail here. Great powers can exercise monopsony-like influence in forum shopping, and the supply of both computing power and machine learning expertise are highly concentrated. In sum, centralisation can reduce competition and enhance coordination, but it may suffocate the creative self-organisation of more fragmented arrangements over time. 

\section{Discussion: What Would History Suggest?}
\subsection{A Summary of Considerations}

The multilateral track record and peculiarities of AI yield suggestions and warnings for the future. A centralised regime could lower costs, support participation, and act as a powerful new linchpin within the international system. Yet centralisation presents risks for AI governance. It could simply produce a brittle dinosaur, of symbolic value but with little meaningful impact on underlying political or technological issues. A poorly executed attempt could lock-in a poorly designed centralised body: a fate worse than fragmentation. Accordingly, ongoing efforts at the UN, OECD, and elsewhere could benefit from addressing the considerations presented in this paper, a summary of which is presented in Appendix A.

\subsection{The Limitations of `Centralisation vs. Decentralisation' Debates}
Structure is not a panacea. Specific provisions such as agendas and decision-making procedures matter greatly, as do the surrounding politics. Underlying political will may be impacted by framing or connecting policy issues \cite[pp.770-1]{koremenos_rational_2001}. The success of a regime is not just a result of fragmentation, but of design details. 

Moreover, institutions can be dynamic and broaden over time by taking in new members, or deepen in strengthening commitments. Successful multilateral efforts, such as trade and ozone depletion, tend to do both. We are in the early days of global AI governance. Decisions taken early on will constrain and partially determine the future path. This dependency can even take place across regimes. The Kyoto Protocol was largely shaped by the targets and timetables approach of the Montreal Protocol, which in turn drew from the Convention on Long-range Transboundary Air Pollution. This targets and timetables approach continues today in the way that most countries frame their climate pledges to the Paris Agreement. The choices we make on governing short-term AI challenges will likely shape the management of other policy issues in the long term \cite{cave_bridging_2019}.

On the other hand, committing to centralisation, even if successful, may amount to solving the wrong problem. The problem may not be structural, but geopolitical. Centralisation could even exacerbate the problem by diluting scarce political attention, incurring heavy transaction costs, and shifting discussions away from bodies which have accumulated experience and practice \cite{juma_commentary:_2000}. For example, the Bretton Woods Institutions of the IMF and World Bank, joined later by the WTO, are centralised regimes that engender power. However, those institutions had the express support of the US and may have simply manifested state power in institutional form. Efforts to ban LAWS and create a cyberwarfare convention have been broadly opposed by states with an established technological superiority in these areas \cite{eilstrup-sangiovanni_why_2018}. A centralised regime may not unpick these power struggles, but just add a layer of complexity.

\begin{table*}[!htbp]
\renewcommand{\arraystretch}{1.5}
\centering
\caption{Regime Complex Monitoring Suggestions}
\resizebox{1\textwidth}{!}{ 
    \begin{tabular}{|l|l|l|}\hline
  
    {\bf Key theme} &
    {\bf Question} &
    {\bf Methods}\\ 
    \hline
    
    Conflict & To what extent are regimes' principles and outputs in opposition over time? & \multirow{3}{3.5in}{Expert and practitioner survey \\ Network analysis (e.g, citation network clustering and centrality) \\ Natural Language Processing (e.g., entailment and fact checking)} \\
    \cline{1-2} 
    
    Coordination & Are regimes taking steps to complement each other? &\\
    \cline{1-2}

    Catalyst & Is the regime complex self-organizing to proactively fill governance gaps?&\\ 
    \cline{1-2}
 
    \hline
    \end{tabular}
}
\label{table1}
\end{table*}

\section{Lessons and Conclusions}

Our framework provides a tool for policy-makers to inform their decisions of whether to join, create, or forgo new institutions that tackle AI policy problems. For instance, the recent choice of whether to support the creation of an independent IPAI involved these considerations. Following the US veto, ongoing negotiations for its replacement at the OECD may similarly benefit from their consideration. For now, it is worth closely monitoring the current landscape of AI governance to see if it exhibits enough policy coordination and political power to effectively deal with mounting AI policy problems. While there are promising initial signs \cite{kunz_artificial_2020} there are also already growing governance failures in LAWS, cyberwarfare, and elsewhere.

We outline a suggested monitoring method in Table 1. There are three key areas to monitor: conflict, coordination, and catalyst. First, {\it conflict} should measure the extent to which principles, rules, regulations and other outcomes from different bodies in the AI regime complex undermine or contradict each other or are in tension either in their principles or goals. Second, {\it coordination} seeks to measure the proactive steps that AI-related regimes take to work with each other. This includes liaison relationships, joint initiatives, as well as the extent to which their rules, outputs and principles tend to reinforce one another. Third, {\it catalyst} raises the important question of governance gaps: is the regime complex self-organising to proactively address international AI policy problems? Numerous AI policy problems currently have no clear coverage under international law, including AI-enabled cyber warfare and HLMI. Whether this changes is of vital importance. 

These areas require investigation through multiple methods. Qualitative surveys of relevant organisations and actors can yield data on expert perceptions of these questions. Surveys can be augmented with quantitative methods, including network analyses of the regime complex relations \cite[p.32]{orsini_regime_2013}. Natural language processing could be used to examine contradictions and similarities between different regime outputs, e.g., statements, meeting minutes, and more. Monitoring the outcomes of fragmentation can help to determine whether centralisation is needed. One way forward would be to empower the OECD AI Policy Observatory or the UN CEB to regularly review the monitoring outcomes. This could inform a democratic discussion and decision of whether to centralise AI governance further. 

Our framework and discussion may also be useful for non-state actors. Researchers and leading AI firms can play an important role in sharing technical expertise and informing forecasts of new policy problems on the horizon. The considerations may benefit their decisions of where to engage. Civil society has a key role as participants, watch-dogs, and catalysts. For example, the Campaign to Stop Killer Robots has sought to boost engagement and support for a LAWS ban within the CCW. Given prolonged delays and a pessimistic outlook, some have articulated a strategy of creating an entirely new forum for the ban, inspired by the Ottawa Treaty which outlawed landmines. Our framework can help reveal the potential virtues (allowing for progress while avoiding high-threshold deadlocks) and vices (enabling forum shopping) of such an approach. It could even help inform the structure of a future international institution, such as allowing for a modular, flexible structure with `critical mass' agreements. One cross-cutting consideration is clear: a fractured regime sees higher participation costs that may threaten to exclude many civil society organisations altogether.

The international governance of AI is nascent and fragmented. Centralisation under a well-designed, modular, `innovation-proof' framework organisation may be a desirable solution. However, such a move must be approached with caution. How to define its scope and mandate is one problem. Ensuring a politically-acceptable and well-designed body is perhaps a more daunting one. It risks cementing in place a fate worse than fragmentation. Monitoring conflict and coordination in the current AI regime complex, and whether governance gaps are filled, is a prudent way of knowing whether the existing structure can suffice. For now we should closely watch the trajectory of both AI technology and its governance initiatives to determine whether centralisation is worth the risk.

\begin{acks}
The authors would like to express thanks to Seth Baum, Haydn Belfield, Jessica Cussins-Newman, Martina Kunz, Jade Leung, Nicolas Mo{\"e}s, Robert de Neufville, and Nicolas Zahn for valuable comments. Any remaining errors are our own. No conflict of interest is identified.
\end{acks}

%
\bibliographystyle{ACM-Reference-Format}
\bibliography{references}


\begin{thebibliography}{50}


\ifx \showCODEN    \undefined \def \showCODEN     #1{\unskip}     \fi
\ifx \showDOI      \undefined \def \showDOI       #1{#1}\fi
\ifx \showISBNx    \undefined \def \showISBNx     #1{\unskip}     \fi
\ifx \showISBNxiii \undefined \def \showISBNxiii  #1{\unskip}     \fi
\ifx \showISSN     \undefined \def \showISSN      #1{\unskip}     \fi
\ifx \showLCCN     \undefined \def \showLCCN      #1{\unskip}     \fi
\ifx \shownote     \undefined \def \shownote      #1{#1}          \fi
\ifx \showarticletitle \undefined \def \showarticletitle #1{#1}   \fi
\ifx \showURL      \undefined \def \showURL       {\relax}        \fi
\providecommand\bibfield[2]{#2}
\providecommand\bibinfo[2]{#2}
\providecommand\natexlab[1]{#1}
\providecommand\showeprint[2][]{arXiv:#2}

\bibitem[\protect\citeauthoryear{Anton}{Anton}{2012}]%
        {anton_treaty_2012}
\bibfield{author}{\bibinfo{person}{Don Anton}.}
  \bibinfo{year}{2012}\natexlab{}.
\newblock \showarticletitle{'{Treaty} {Congestion}' in {International}
  {Environmental} {Law}}.
\newblock In \bibinfo{booktitle}{\emph{Routledge {Handbook} of {International}
  {Environmental} {Law}}}, \bibfield{editor}{\bibinfo{person}{Shawkat Alam},
  \bibinfo{person}{Jahid~Hossain Bhuiyan}, \bibinfo{person}{Tareq~M.R.
  Chowdhury}, {and} \bibinfo{person}{Erika~J. Techera}} (Eds.).
  \bibinfo{publisher}{Routledge}, \bibinfo{address}{London}.
\newblock
\urldef\tempurl%
\url{https://www.taylorfrancis.com/books/9780203093474}
\showURL{%
\tempurl}


\bibitem[\protect\citeauthoryear{Baccaro and Mele}{Baccaro and Mele}{2012}]%
        {baccaro_pathology_2012}
\bibfield{author}{\bibinfo{person}{Lucio Baccaro} {and}
  \bibinfo{person}{Valentina Mele}.} \bibinfo{year}{2012}\natexlab{}.
\newblock \showarticletitle{Pathology of {Path} {Dependency}? {The} {ILO} and
  the {Challenge} of {New} {Governance}}.
\newblock \bibinfo{journal}{\emph{ILR Review}} \bibinfo{volume}{65},
  \bibinfo{number}{2} (\bibinfo{date}{April} \bibinfo{year}{2012}),
  \bibinfo{pages}{195--224}.
\newblock
\showISSN{0019-7939, 2162-271X}
\urldef\tempurl%
\url{https://doi.org/10.1177/001979391206500201}
\showDOI{\tempurl}


\bibitem[\protect\citeauthoryear{Biermann, Pattberg, van Asselt, and
  Zelli}{Biermann et~al\mbox{.}}{2009}]%
        {biermann_fragmentation_2009}
\bibfield{author}{\bibinfo{person}{Frank Biermann}, \bibinfo{person}{Philipp
  Pattberg}, \bibinfo{person}{Harro van Asselt}, {and}
  \bibinfo{person}{Fariborz Zelli}.} \bibinfo{year}{2009}\natexlab{}.
\newblock \showarticletitle{The {Fragmentation} of {Global} {Governance}
  {Architectures}: {A} {Framework} for {Analysis}}.
\newblock \bibinfo{journal}{\emph{Global Environmental Politics}}
  \bibinfo{volume}{9}, \bibinfo{number}{4} (\bibinfo{date}{Oct.}
  \bibinfo{year}{2009}), \bibinfo{pages}{14--40}.
\newblock
\showISSN{1526-3800}
\urldef\tempurl%
\url{https://doi.org/10.1162/glep.2009.9.4.14}
\showDOI{\tempurl}


\bibitem[\protect\citeauthoryear{Braithwaite and Drahos}{Braithwaite and
  Drahos}{2000}]%
        {braithwaite_global_2000}
\bibfield{author}{\bibinfo{person}{John Braithwaite} {and}
  \bibinfo{person}{Peter Drahos}.} \bibinfo{year}{2000}\natexlab{}.
\newblock \bibinfo{booktitle}{\emph{Global {Business} {Regulation}}}.
\newblock \bibinfo{publisher}{Cambridge University Press},
  \bibinfo{address}{Cambridge}.
\newblock
\showISBNx{978-0-521-78499-3}
\newblock
\shownote{Google-Books-ID: DcEEW5OGWLcC.}


\bibitem[\protect\citeauthoryear{Butcher and Beridze}{Butcher and
  Beridze}{2019}]%
        {UNIRI2019}
\bibfield{author}{\bibinfo{person}{James Butcher} {and} \bibinfo{person}{Irakli
  Beridze}.} \bibinfo{year}{2019}\natexlab{}.
\newblock \showarticletitle{What is the State of Artificial Intelligence
  Governance Globally?}
\newblock \bibinfo{journal}{\emph{The RUSI Journal}} \bibinfo{volume}{164},
  \bibinfo{number}{5-6} (\bibinfo{year}{2019}), \bibinfo{pages}{88--96}.
\newblock
\urldef\tempurl%
\url{https://doi.org/10.1080/03071847.2019.1694260}
\showDOI{\tempurl}
\showeprint{https://doi.org/10.1080/03071847.2019.1694260}


\bibitem[\protect\citeauthoryear{Calo}{Calo}{2017}]%
        {calo_artificial_2017}
\bibfield{author}{\bibinfo{person}{Ryan Calo}.}
  \bibinfo{year}{2017}\natexlab{}.
\newblock \showarticletitle{Artificial {Intelligence} {Policy}: {A} {Primer}
  and {Roadmap}}.
\newblock \bibinfo{journal}{\emph{UC Davis Law Review}}  \bibinfo{volume}{51}
  (\bibinfo{year}{2017}), \bibinfo{pages}{37}.
\newblock
\urldef\tempurl%
\url{https://lawreview.law.ucdavis.edu/issues/51/2/Symposium/51-2_Calo.pdf}
\showURL{%
\tempurl}


\bibitem[\protect\citeauthoryear{Cave and \'{O}h\'{E}igeartaigh}{Cave and
  \'{O}h\'{E}igeartaigh}{2019}]%
        {cave_bridging_2019}
\bibfield{author}{\bibinfo{person}{Stephen Cave} {and}
  \bibinfo{person}{Se\'{a}n~S. \'{O}h\'{E}igeartaigh}.}
  \bibinfo{year}{2019}\natexlab{}.
\newblock \showarticletitle{Bridging near- and long-term concerns about {AI}}.
\newblock \bibinfo{journal}{\emph{Nature Machine Intelligence}}
  \bibinfo{volume}{1}, \bibinfo{number}{1} (\bibinfo{date}{Jan.}
  \bibinfo{year}{2019}), \bibinfo{pages}{5}.
\newblock
\showISSN{2522-5839}
\urldef\tempurl%
\url{https://doi.org/10.1038/s42256-018-0003-2}
\showDOI{\tempurl}


\bibitem[\protect\citeauthoryear{Cihon}{Cihon}{2019}]%
        {cihon_standards_2019}
\bibfield{author}{\bibinfo{person}{Peter Cihon}.}
  \bibinfo{year}{2019}\natexlab{}.
\newblock \bibinfo{booktitle}{\emph{Standards for {AI} {Governance}:
  {International} {Standards} to {Enable} {Global} {Coordination} in {AI}
  {Research} \& {Development}}}.
\newblock \bibinfo{type}{Technical {Report}}. \bibinfo{institution}{Center for
  the Governance of AI, Future of Humanity Institute, University of Oxford},
  \bibinfo{address}{Oxford}.
\newblock
\urldef\tempurl%
\url{https://www.fhi.ox.ac.uk/wp-content/uploads/Standards_-FHI-Technical-Report.pdf}
\showURL{%
\tempurl}


\bibitem[\protect\citeauthoryear{Dafoe}{Dafoe}{2018}]%
        {dafoe_ai_2018}
\bibfield{author}{\bibinfo{person}{Allan Dafoe}.}
  \bibinfo{year}{2018}\natexlab{}.
\newblock \bibinfo{booktitle}{\emph{{AI} {Governance}: {A} {Research}
  {Agenda}}}.
\newblock \bibinfo{type}{{T}echnical {R}eport}. \bibinfo{institution}{Center
  for the Governance of AI, Future of Humanity Institute},
  \bibinfo{address}{Oxford}. \bibinfo{pages}{52} pages.
\newblock
\urldef\tempurl%
\url{https://www.fhi.ox.ac.uk/govaiagenda/}
\showURL{%
\tempurl}


\bibitem[\protect\citeauthoryear{Delcker}{Delcker}{2019}]%
        {delcker_how_2019}
\bibfield{author}{\bibinfo{person}{Janosch Delcker}.}
  \bibinfo{year}{2019}\natexlab{}.
\newblock \showarticletitle{How killer robots overran the {UN}}.
\newblock \bibinfo{journal}{\emph{POLITICO}} (\bibinfo{date}{Feb.}
  \bibinfo{year}{2019}).
\newblock
\urldef\tempurl%
\url{https://www.politico.eu/article/killer-robots-overran-united-nations-lethal-autonomous-weapons-systems/}
\showURL{%
\tempurl}


\bibitem[\protect\citeauthoryear{Deshman}{Deshman}{2011}]%
        {deshman_horizontal_2011}
\bibfield{author}{\bibinfo{person}{Abigail~C. Deshman}.}
  \bibinfo{year}{2011}\natexlab{}.
\newblock \showarticletitle{Horizontal {Review} between {International}
  {Organizations}: {Why}, {How}, and {Who} {Cares} about {Corporate}
  {Regulatory} {Capture}}.
\newblock \bibinfo{journal}{\emph{European Journal of International Law}}
  \bibinfo{volume}{22}, \bibinfo{number}{4} (\bibinfo{date}{Nov.}
  \bibinfo{year}{2011}), \bibinfo{pages}{1089--1113}.
\newblock
\showISSN{0938-5428}
\urldef\tempurl%
\url{https://doi.org/10.1093/ejil/chr093}
\showDOI{\tempurl}


\bibitem[\protect\citeauthoryear{Eckersley}{Eckersley}{2004}]%
        {eckersley_big_2004}
\bibfield{author}{\bibinfo{person}{Robyn Eckersley}.}
  \bibinfo{year}{2004}\natexlab{}.
\newblock \showarticletitle{The {Big} {Chill}: {The} {WTO} and {Multilateral}
  {Environmental} {Agreements}}.
\newblock \bibinfo{journal}{\emph{Global Environmental Politics}}
  \bibinfo{volume}{4}, \bibinfo{number}{2} (\bibinfo{date}{May}
  \bibinfo{year}{2004}), \bibinfo{pages}{24--50}.
\newblock
\showISSN{1526-3800, 1536-0091}
\urldef\tempurl%
\url{https://doi.org/10.1162/152638004323074183}
\showDOI{\tempurl}


\bibitem[\protect\citeauthoryear{Eilstrup-Sangiovanni}{Eilstrup-Sangiovanni}{2018}]%
        {eilstrup-sangiovanni_why_2018}
\bibfield{author}{\bibinfo{person}{Mette Eilstrup-Sangiovanni}.}
  \bibinfo{year}{2018}\natexlab{}.
\newblock \showarticletitle{Why the {World} {Needs} an {International}
  {Cyberwar} {Convention}}.
\newblock \bibinfo{journal}{\emph{Philosophy \& Technology}}
  \bibinfo{volume}{31}, \bibinfo{number}{3} (\bibinfo{date}{Sept.}
  \bibinfo{year}{2018}), \bibinfo{pages}{379--407}.
\newblock
\showISSN{2210-5433, 2210-5441}
\urldef\tempurl%
\url{https://doi.org/10.1007/s13347-017-0271-5}
\showDOI{\tempurl}


\bibitem[\protect\citeauthoryear{Erdelyi and Goldsmith}{Erdelyi and
  Goldsmith}{2018}]%
        {erdelyi_regulating_2018}
\bibfield{author}{\bibinfo{person}{Olivia~J Erdelyi} {and}
  \bibinfo{person}{Judy Goldsmith}.} \bibinfo{year}{2018}\natexlab{}.
\newblock \showarticletitle{Regulating {Artificial} {Intelligence}: {Proposal}
  for a {Global} {Solution}}. In \bibinfo{booktitle}{\emph{Proceedings of the
  2018 {AAAI} / {ACM} {Conference} on {Artificial} {Intelligence}, {Ethics} and
  {Society}}}. \bibinfo{publisher}{AAAI}, \bibinfo{address}{Palo Alto, CA},
  \bibinfo{pages}{7}.
\newblock
\urldef\tempurl%
\url{https://par.nsf.gov/servlets/purl/10066933}
\showURL{%
\tempurl}


\bibitem[\protect\citeauthoryear{Esty and Ivanova}{Esty and Ivanova}{2002}]%
        {esty_revitalizing_2002}
\bibfield{author}{\bibinfo{person}{Daniel~C Esty} {and}
  \bibinfo{person}{Maria~H Ivanova}.} \bibinfo{year}{2002}\natexlab{}.
\newblock \showarticletitle{Revitalizing {Global} {Environmental} {Governance}:
  {A} {Function}-{Driven} {Approach}}.
\newblock In \bibinfo{booktitle}{\emph{Global {Environmental} {Governance}:
  {Options} \& {Opportunities}}}, \bibfield{editor}{\bibinfo{person}{Daniel~C
  Esty} {and} \bibinfo{person}{Maria~H Ivanova}} (Eds.).
  \bibinfo{publisher}{Yale School of Forestry and Environmental Studies},
  \bibinfo{address}{New Haven, CT}.
\newblock
\urldef\tempurl%
\url{https://environment.yale.edu/publication-series/documents/downloads/a-g/esty-ivanova.pdf}
\showURL{%
\tempurl}


\bibitem[\protect\citeauthoryear{Gasser and Almeida}{Gasser and
  Almeida}{2017}]%
        {gasser_layered_2017}
\bibfield{author}{\bibinfo{person}{Urs Gasser} {and}
  \bibinfo{person}{Virgilio~A.F. Almeida}.} \bibinfo{year}{2017}\natexlab{}.
\newblock \showarticletitle{A {Layered} {Model} for {AI} {Governance}}.
\newblock \bibinfo{journal}{\emph{IEEE Internet Computing}}
  \bibinfo{volume}{21}, \bibinfo{number}{6} (\bibinfo{date}{Nov.}
  \bibinfo{year}{2017}), \bibinfo{pages}{58--62}.
\newblock
\showISSN{1089-7801}
\urldef\tempurl%
\url{https://doi.org/10.1109/MIC.2017.4180835}
\showDOI{\tempurl}


\bibitem[\protect\citeauthoryear{Grace, Salvatier, Dafoe, Zhang, and
  Evans}{Grace et~al\mbox{.}}{2018}]%
        {grace2018will}
\bibfield{author}{\bibinfo{person}{Katja Grace}, \bibinfo{person}{John
  Salvatier}, \bibinfo{person}{Allan Dafoe}, \bibinfo{person}{Baobao Zhang},
  {and} \bibinfo{person}{Owain Evans}.} \bibinfo{year}{2018}\natexlab{}.
\newblock \showarticletitle{When will AI exceed human performance? Evidence
  from AI experts}.
\newblock \bibinfo{journal}{\emph{Journal of Artificial Intelligence Research}}
   \bibinfo{volume}{62} (\bibinfo{year}{2018}), \bibinfo{pages}{729--754}.
\newblock


\bibitem[\protect\citeauthoryear{Helfer}{Helfer}{2004}]%
        {helfer_regime_2004}
\bibfield{author}{\bibinfo{person}{Laurence Helfer}.}
  \bibinfo{year}{2004}\natexlab{}.
\newblock \showarticletitle{Regime {Shifting}: {The} {TRIPs} {Agreement} and
  {New} {Dynamics} of {International} {Intellectual} {Property} {Lawmaking}}.
\newblock \bibinfo{journal}{\emph{Yale Journal of International Law}}
  \bibinfo{volume}{29} (\bibinfo{date}{Jan.} \bibinfo{year}{2004}),
  \bibinfo{pages}{1--83}.
\newblock
\urldef\tempurl%
\url{https://scholarship.law.duke.edu/faculty_scholarship/2014}
\showURL{%
\tempurl}


\bibitem[\protect\citeauthoryear{High-Level~Panel}{High-Level~Panel}{2019}]%
        {high2019panel}
\bibfield{author}{\bibinfo{person}{on~Digital~Cooperation High-Level~Panel}.}
  \bibinfo{year}{2019}\natexlab{}.
\newblock \showarticletitle{The Age of Digital Interdependence Report}.
\newblock \bibinfo{journal}{\emph{UN Secretary General}}
  (\bibinfo{year}{2019}).
\newblock


\bibitem[\protect\citeauthoryear{ITU}{ITU}{2019}]%
        {itu_united_2019}
\bibfield{author}{\bibinfo{person}{ITU}.} \bibinfo{year}{2019}\natexlab{}.
\newblock \bibinfo{booktitle}{\emph{United {Nations} {Activities} on
  {Artificial} {Intelligence} ({AI}) 2019}}.
\newblock \bibinfo{type}{{T}echnical {R}eport}. \bibinfo{institution}{ITU}.
  \bibinfo{pages}{88} pages.
\newblock
\urldef\tempurl%
\url{https://www.itu.int/dms_pub/itu-s/opb/gen/S-GEN-UNACT-2019-1-PDF-E.pdf}
\showURL{%
\tempurl}


\bibitem[\protect\citeauthoryear{Juma}{Juma}{2000}]%
        {juma_commentary:_2000}
\bibfield{author}{\bibinfo{person}{Calestous Juma}.}
  \bibinfo{year}{2000}\natexlab{}.
\newblock \showarticletitle{Commentary: {The} {Perils} of {Centralizing}
  {Global} {Environmental} {Governance}}.
\newblock \bibinfo{journal}{\emph{Environment: Science and Policy for
  Sustainable Development}} \bibinfo{volume}{42}, \bibinfo{number}{9}
  (\bibinfo{date}{Nov.} \bibinfo{year}{2000}), \bibinfo{pages}{44--45}.
\newblock
\showISSN{0013-9157}
\urldef\tempurl%
\url{https://doi.org/10.1080/00139150009605765}
\showDOI{\tempurl}


\bibitem[\protect\citeauthoryear{Jupille, Mattli, and Snidal}{Jupille
  et~al\mbox{.}}{2013}]%
        {jupille_institutional_2013}
\bibfield{author}{\bibinfo{person}{Joseph Jupille}, \bibinfo{person}{Walter
  Mattli}, {and} \bibinfo{person}{Duncan Snidal}.}
  \bibinfo{year}{2013}\natexlab{}.
\newblock \bibinfo{booktitle}{\emph{Institutional {Choice} and {Global}
  {Commerce}}}.
\newblock \bibinfo{publisher}{Cambridge University Press},
  \bibinfo{address}{Cambridge}.
\newblock
\showISBNx{978-1-107-03895-0 978-1-107-64592-9 978-1-139-85599-0}
\newblock
\shownote{OCLC: 900490808.}


\bibitem[\protect\citeauthoryear{Kemp}{Kemp}{2017}]%
        {kemp_us-proofing_2017}
\bibfield{author}{\bibinfo{person}{Luke Kemp}.}
  \bibinfo{year}{2017}\natexlab{}.
\newblock \showarticletitle{{US}-proofing the {Paris} {Climate} {Agreement}}.
\newblock \bibinfo{journal}{\emph{Climate Policy}} \bibinfo{volume}{17},
  \bibinfo{number}{1} (\bibinfo{date}{Jan.} \bibinfo{year}{2017}),
  \bibinfo{pages}{86--101}.
\newblock
\showISSN{1469-3062}
\urldef\tempurl%
\url{https://doi.org/10.1080/14693062.2016.1176007}
\showDOI{\tempurl}


\bibitem[\protect\citeauthoryear{Kemp, Cihon, Maas, Belfield, Cremer, Leung,
  and Ó~hÉigeartaigh}{Kemp et~al\mbox{.}}{2019}]%
        {kemp_high-level_2019}
\bibfield{author}{\bibinfo{person}{Luke Kemp}, \bibinfo{person}{Peter Cihon},
  \bibinfo{person}{Matthijs~Michiel Maas}, \bibinfo{person}{Haydn Belfield},
  \bibinfo{person}{Zoe Cremer}, \bibinfo{person}{Jade Leung}, {and}
  \bibinfo{person}{Seán Ó~hÉigeartaigh}.} \bibinfo{year}{2019}\natexlab{}.
\newblock \bibinfo{title}{{UN} {High}-level {Panel} on {Digital} {Cooperation}:
  {A} {Proposal} for {International} {AI} {Governance}}.
\newblock
\newblock
\urldef\tempurl%
\url{https://digitalcooperation.org/wp-content/uploads/2019/02/Luke_Kemp_Submission-to-the-UN-High-Level-Panel-on-Digital-Cooperation-2019-Kemp-et-al.pdf}
\showURL{%
\tempurl}


\bibitem[\protect\citeauthoryear{Koremenos, Lipson, and Snidal}{Koremenos
  et~al\mbox{.}}{2001a}]%
        {koremenos_rational_2001-1}
\bibfield{author}{\bibinfo{person}{Barbara Koremenos}, \bibinfo{person}{Charles
  Lipson}, {and} \bibinfo{person}{Duncan Snidal}.}
  \bibinfo{year}{2001}\natexlab{a}.
\newblock \showarticletitle{Rational {Design}: {Looking} {Back} to {Move}
  {Forward}}.
\newblock \bibinfo{journal}{\emph{International Organization}}
  \bibinfo{volume}{55}, \bibinfo{number}{4} (\bibinfo{year}{2001}),
  \bibinfo{pages}{1051--1082}.
\newblock
\showISSN{1531-5088, 0020-8183}
\urldef\tempurl%
\url{https://doi.org/10.1162/002081801317193691}
\showDOI{\tempurl}


\bibitem[\protect\citeauthoryear{Koremenos, Lipson, and Snidal}{Koremenos
  et~al\mbox{.}}{2001b}]%
        {koremenos_rational_2001}
\bibfield{author}{\bibinfo{person}{Barbara Koremenos}, \bibinfo{person}{Charles
  Lipson}, {and} \bibinfo{person}{Duncan Snidal}.}
  \bibinfo{year}{2001}\natexlab{b}.
\newblock \showarticletitle{The {Rational} {Design} of {International}
  {Institutions}}.
\newblock \bibinfo{journal}{\emph{International Organization}}
  \bibinfo{volume}{55}, \bibinfo{number}{4} (\bibinfo{year}{2001}),
  \bibinfo{pages}{761--799}.
\newblock
\showISSN{1531-5088, 0020-8183}
\urldef\tempurl%
\url{https://doi.org/10.1162/002081801317193592}
\showDOI{\tempurl}


\bibitem[\protect\citeauthoryear{Krasner}{Krasner}{1982}]%
        {krasner_structural_1982}
\bibfield{author}{\bibinfo{person}{Stephen~D. Krasner}.}
  \bibinfo{year}{1982}\natexlab{}.
\newblock \showarticletitle{Structural {Causes} and {Regime} {Consequences}:
  {Regimes} as {Intervening} {Variables}}.
\newblock \bibinfo{journal}{\emph{International Organization}}
  \bibinfo{volume}{36}, \bibinfo{number}{2} (\bibinfo{year}{1982}),
  \bibinfo{pages}{185--205}.
\newblock
\showISSN{0020-8183}
\urldef\tempurl%
\url{https://doi.org/10.1017/S0020818300018920}
\showDOI{\tempurl}


\bibitem[\protect\citeauthoryear{Kunz and \'{O}h\'{E}igeartaigh}{Kunz and
  \'{O}h\'{E}igeartaigh}{2020}]%
        {kunz_artificial_2020}
\bibfield{author}{\bibinfo{person}{Martina Kunz} {and}
  \bibinfo{person}{Se\'{a}n \'{O}h\'{E}igeartaigh}.}
  \bibinfo{year}{2020}\natexlab{}.
\newblock \showarticletitle{Artificial {Intelligence} and {Robotization}}.
\newblock In \bibinfo{booktitle}{\emph{Oxford {Handbook} on the {International}
  {Law} of {Global} {Security}}}, \bibfield{editor}{\bibinfo{person}{Robin
  Geiss} {and} \bibinfo{person}{Nils Melzer}} (Eds.).
  \bibinfo{publisher}{Oxford University Press}, \bibinfo{address}{Oxford}.
\newblock
\urldef\tempurl%
\url{https://papers.ssrn.com/abstract=3310421}
\showURL{%
\tempurl}


\bibitem[\protect\citeauthoryear{Maas}{Maas}{2019}]%
        {maas_innovation-proof_2019}
\bibfield{author}{\bibinfo{person}{Matthijs~M. Maas}.}
  \bibinfo{year}{2019}\natexlab{}.
\newblock \showarticletitle{Innovation-{Proof} {Governance} for {Military}
  {AI}? how {I} learned to stop worrying and love the bot}.
\newblock \bibinfo{journal}{\emph{Journal of International Humanitarian Legal
  Studies}} \bibinfo{volume}{10}, \bibinfo{number}{1} (\bibinfo{year}{2019}),
  \bibinfo{pages}{129--157}.
\newblock
\urldef\tempurl%
\url{https://doi.org/10.1163/18781527-01001006}
\showDOI{\tempurl}


\bibitem[\protect\citeauthoryear{Marchant, Allenby, and Herkert}{Marchant
  et~al\mbox{.}}{2011}]%
        {marchant2011growing}
\bibfield{author}{\bibinfo{person}{Gary~E Marchant}, \bibinfo{person}{Braden~R
  Allenby}, {and} \bibinfo{person}{Joseph~R Herkert}.}
  \bibinfo{year}{2011}\natexlab{}.
\newblock \bibinfo{booktitle}{\emph{The growing gap between emerging
  technologies and legal-ethical oversight: The pacing problem}}.
  Vol.~\bibinfo{volume}{7}.
\newblock \bibinfo{publisher}{Springer Science \& Business Media},
  \bibinfo{address}{Berlin}.
\newblock


\bibitem[\protect\citeauthoryear{Martin and Messerlin}{Martin and
  Messerlin}{2007}]%
        {martin2007so}
\bibfield{author}{\bibinfo{person}{Will Martin} {and} \bibinfo{person}{Patrick
  Messerlin}.} \bibinfo{year}{2007}\natexlab{}.
\newblock \showarticletitle{Why is it so difficult? Trade liberalization under
  the Doha Agenda}.
\newblock \bibinfo{journal}{\emph{Oxford Review of Economic Policy}}
  \bibinfo{volume}{23}, \bibinfo{number}{3} (\bibinfo{year}{2007}),
  \bibinfo{pages}{347--366}.
\newblock


\bibitem[\protect\citeauthoryear{Morin, Dobson, Peacock, Prys‐Hansen, Anne,
  Bélanger, Dietsch, Fabian, Kirton, Marchetti, Romano, Schreurs, Silve, and
  Vallet}{Morin et~al\mbox{.}}{2019}]%
        {morin_how_2019}
\bibfield{author}{\bibinfo{person}{Jean‐Frédéric Morin},
  \bibinfo{person}{Hugo Dobson}, \bibinfo{person}{Claire Peacock},
  \bibinfo{person}{Miriam Prys‐Hansen}, \bibinfo{person}{Abdoulaye Anne},
  \bibinfo{person}{Louis Bélanger}, \bibinfo{person}{Peter Dietsch},
  \bibinfo{person}{Judit Fabian}, \bibinfo{person}{John Kirton},
  \bibinfo{person}{Raffaele Marchetti}, \bibinfo{person}{Simone Romano},
  \bibinfo{person}{Miranda Schreurs}, \bibinfo{person}{Arthur Silve}, {and}
  \bibinfo{person}{Elisabeth Vallet}.} \bibinfo{year}{2019}\natexlab{}.
\newblock \showarticletitle{How {Informality} {Can} {Address} {Emerging}
  {Issues}: {Making} the {Most} of the {G}7}.
\newblock \bibinfo{journal}{\emph{Global Policy}} \bibinfo{volume}{10},
  \bibinfo{number}{2} (\bibinfo{date}{May} \bibinfo{year}{2019}),
  \bibinfo{pages}{267--273}.
\newblock
\showISSN{1758-5880, 1758-5899}
\urldef\tempurl%
\url{https://doi.org/10.1111/1758-5899.12668}
\showDOI{\tempurl}


\bibitem[\protect\citeauthoryear{Mueller, Mathiason, and Klein}{Mueller
  et~al\mbox{.}}{2007}]%
        {mueller_internet_2007}
\bibfield{author}{\bibinfo{person}{Milton Mueller}, \bibinfo{person}{John
  Mathiason}, {and} \bibinfo{person}{Hans Klein}.}
  \bibinfo{year}{2007}\natexlab{}.
\newblock \showarticletitle{The {Internet} and {Global} {Governance}:
  {Principles} and {Norms} for a {New} {Regime}}.
\newblock \bibinfo{journal}{\emph{Global Governance}} \bibinfo{volume}{13},
  \bibinfo{number}{2} (\bibinfo{year}{2007}), \bibinfo{pages}{237--254}.
\newblock
\urldef\tempurl%
\url{https://heinonline.org/HOL/P?h=hein.journals/glogo13&i=245}
\showURL{%
\tempurl}


\bibitem[\protect\citeauthoryear{Nilsson}{Nilsson}{2009}]%
        {nilsson_quest_2009}
\bibfield{author}{\bibinfo{person}{Nils~J. Nilsson}.}
  \bibinfo{year}{2009}\natexlab{}.
\newblock \bibinfo{booktitle}{\emph{The {Quest} for {Artificial}
  {Intelligence}} (\bibinfo{edition}{1 edition} ed.)}.
\newblock \bibinfo{publisher}{Cambridge University Press},
  \bibinfo{address}{Cambridge ; New York}.
\newblock
\showISBNx{978-0-521-12293-1}


\bibitem[\protect\citeauthoryear{Nye}{Nye}{2014}]%
        {nye_regime_2014}
\bibfield{author}{\bibinfo{person}{Joseph~S. Nye}.}
  \bibinfo{year}{2014}\natexlab{}.
\newblock \bibinfo{booktitle}{\emph{The {Regime} {Complex} for {Managing}
  {Global} {Cyber} {Activities}}}.
\newblock \bibinfo{type}{{T}echnical {R}eport}~1. \bibinfo{institution}{Global
  Commission on Internet Governance}.
\newblock
\urldef\tempurl%
\url{https://dash.harvard.edu/bitstream/handle/1/12308565/Nye-GlobalCommission.pdf}
\showURL{%
\tempurl}


\bibitem[\protect\citeauthoryear{Orsini, Morin, and Young}{Orsini
  et~al\mbox{.}}{2013}]%
        {orsini_regime_2013}
\bibfield{author}{\bibinfo{person}{Amandine Orsini},
  \bibinfo{person}{Jean-Fr\'{e}d\'{e}ric Morin}, {and} \bibinfo{person}{Oran
  Young}.} \bibinfo{year}{2013}\natexlab{}.
\newblock \showarticletitle{Regime {Complexes}: {A} {Buzz}, a {Boom}, or a
  {Boost} for {Global} {Governance}?}
\newblock \bibinfo{journal}{\emph{Global Governance: A Review of
  Multilateralism and International Organizations}} \bibinfo{volume}{19},
  \bibinfo{number}{1} (\bibinfo{date}{Aug.} \bibinfo{year}{2013}),
  \bibinfo{pages}{27--39}.
\newblock
\showISSN{1942-6720, 1075-2846}
\urldef\tempurl%
\url{https://doi.org/10.1163/19426720-01901003}
\showDOI{\tempurl}


\bibitem[\protect\citeauthoryear{Ostrom}{Ostrom}{2010}]%
        {ostrom_polycentric_2010}
\bibfield{author}{\bibinfo{person}{Elinor Ostrom}.}
  \bibinfo{year}{2010}\natexlab{}.
\newblock \showarticletitle{Polycentric systems for coping with collective
  action and global environmental change}.
\newblock \bibinfo{journal}{\emph{Global Environmental Change}}
  \bibinfo{volume}{20}, \bibinfo{number}{4} (\bibinfo{date}{Oct.}
  \bibinfo{year}{2010}), \bibinfo{pages}{550--557}.
\newblock
\showISSN{09593780}
\urldef\tempurl%
\url{https://doi.org/10.1016/j.gloenvcha.2010.07.004}
\showDOI{\tempurl}


\bibitem[\protect\citeauthoryear{Patrick}{Patrick}{2014}]%
        {patrick_unruled_2014}
\bibfield{author}{\bibinfo{person}{Stewart Patrick}.}
  \bibinfo{year}{2014}\natexlab{}.
\newblock \showarticletitle{The {Unruled} {World}: {The} {Case} for {Good}
  {Enough} {Global} {Governance}}.
\newblock \bibinfo{journal}{\emph{Foreign Affairs}} \bibinfo{volume}{93},
  \bibinfo{number}{1} (\bibinfo{year}{2014}), \bibinfo{pages}{58--73}.
\newblock


\bibitem[\protect\citeauthoryear{Pauwels}{Pauwels}{2019}]%
        {pauwels_new_2019}
\bibfield{author}{\bibinfo{person}{Eleonore Pauwels}.}
  \bibinfo{year}{2019}\natexlab{}.
\newblock \bibinfo{booktitle}{\emph{The {New} {Geopolitics} of {Converging}
  {Risks}: {The} {UN} and {Prevention} in the {Era} of {AI}}}.
\newblock \bibinfo{type}{{T}echnical {R}eport}. \bibinfo{institution}{United
  Nations University - Centre for Policy Research}. \bibinfo{pages}{83} pages.
\newblock
\urldef\tempurl%
\url{https://i.unu.edu/media/cpr.unu.edu/attachment/3472/PauwelsAIGeopolitics.pdf}
\showURL{%
\tempurl}


\bibitem[\protect\citeauthoryear{Pekkanen, Solís, and Katada}{Pekkanen
  et~al\mbox{.}}{2007}]%
        {pekkanen_trading_2007}
\bibfield{author}{\bibinfo{person}{Saadia~M. Pekkanen}, \bibinfo{person}{Mireya
  Solís}, {and} \bibinfo{person}{Saori~N. Katada}.}
  \bibinfo{year}{2007}\natexlab{}.
\newblock \showarticletitle{Trading {Gains} for {Control}: {International}
  {Trade} {Forums} and {Japanese} {Economic} {Diplomacy}}.
\newblock \bibinfo{journal}{\emph{International Studies Quarterly}}
  \bibinfo{volume}{51}, \bibinfo{number}{4} (\bibinfo{year}{2007}),
  \bibinfo{pages}{945--970}.
\newblock
\showISSN{0020-8833}
\urldef\tempurl%
\url{https://www.jstor.org/stable/4621750}
\showURL{%
\tempurl}


\bibitem[\protect\citeauthoryear{Perrault, Shoham, Brynjolfsson, Clark,
  Etchemendy, Grosz, Lyons, Manyika, Mishra, and Niebles}{Perrault
  et~al\mbox{.}}{2019}]%
        {perrault_ai_2019}
\bibfield{author}{\bibinfo{person}{Raymond Perrault}, \bibinfo{person}{Yoav
  Shoham}, \bibinfo{person}{Erik Brynjolfsson}, \bibinfo{person}{Jack Clark},
  \bibinfo{person}{John Etchemendy}, \bibinfo{person}{Barbara Grosz},
  \bibinfo{person}{Terah Lyons}, \bibinfo{person}{James Manyika},
  \bibinfo{person}{Saurabh Mishra}, {and} \bibinfo{person}{Juan~Carlos
  Niebles}.} \bibinfo{year}{2019}\natexlab{}.
\newblock \bibinfo{booktitle}{\emph{The {AI} {Index} 2019 {Annual} {Report}}}.
\newblock \bibinfo{type}{{T}echnical {R}eport}. \bibinfo{institution}{AI Index
  Steering Committee, Human-Centered AI Initiative, Stanford University},
  \bibinfo{address}{Stanford, CA}.
\newblock
\urldef\tempurl%
\url{https://hai.stanford.edu/sites/g/files/sbiybj10986/f/ai_index_2019_report.pdf}
\showURL{%
\tempurl}


\bibitem[\protect\citeauthoryear{Picker}{Picker}{2001}]%
        {picker_view_2001}
\bibfield{author}{\bibinfo{person}{Colin~B. Picker}.}
  \bibinfo{year}{2001}\natexlab{}.
\newblock \showarticletitle{A {View} from 40,000 {Feet}: {International} {Law}
  and the {Invisible} {Hand} of {Technology}}.
\newblock \bibinfo{journal}{\emph{Cardozo Law Review}}  \bibinfo{volume}{23}
  (\bibinfo{year}{2001}), \bibinfo{pages}{151--219}.
\newblock
\urldef\tempurl%
\url{https://papers.ssrn.com/abstract=987524}
\showURL{%
\tempurl}


\bibitem[\protect\citeauthoryear{Rolnick, Donti, Kaack, Kochanski, Lacoste,
  Sankaran, Ross, Milojevic-Dupont, Jaques, Waldman-Brown, Luccioni, Maharaj,
  Sherwin, Mukkavilli, Kording, Gomes, Ng, Hassabis, Platt, Creutzig, Chayes,
  and Bengio}{Rolnick et~al\mbox{.}}{2019}]%
        {rolnick_tackling_2019}
\bibfield{author}{\bibinfo{person}{David Rolnick}, \bibinfo{person}{Priya~L.
  Donti}, \bibinfo{person}{Lynn~H. Kaack}, \bibinfo{person}{Kelly Kochanski},
  \bibinfo{person}{Alexandre Lacoste}, \bibinfo{person}{Kris Sankaran},
  \bibinfo{person}{Andrew~Slavin Ross}, \bibinfo{person}{Nikola
  Milojevic-Dupont}, \bibinfo{person}{Natasha Jaques}, \bibinfo{person}{Anna
  Waldman-Brown}, \bibinfo{person}{Alexandra Luccioni}, \bibinfo{person}{Tegan
  Maharaj}, \bibinfo{person}{Evan~D. Sherwin}, \bibinfo{person}{S.~Karthik
  Mukkavilli}, \bibinfo{person}{Konrad~P. Kording}, \bibinfo{person}{Carla
  Gomes}, \bibinfo{person}{Andrew~Y. Ng}, \bibinfo{person}{Demis Hassabis},
  \bibinfo{person}{John~C. Platt}, \bibinfo{person}{Felix Creutzig},
  \bibinfo{person}{Jennifer Chayes}, {and} \bibinfo{person}{Yoshua Bengio}.}
  \bibinfo{year}{2019}\natexlab{}.
\newblock \bibinfo{title}{Tackling Climate Change with Machine Learning}.
\newblock
\newblock
\showeprint[arxiv]{cs.CY/1906.05433}


\bibitem[\protect\citeauthoryear{Scherer}{Scherer}{2016}]%
        {scherer_regulating_2016}
\bibfield{author}{\bibinfo{person}{Matthew~U. Scherer}.}
  \bibinfo{year}{2016}\natexlab{}.
\newblock \showarticletitle{Regulating {Artificial} {Intelligence} {Systems}:
  {Risks}, {Challenges}, {Competencies}, and {Strategies}}.
\newblock \bibinfo{journal}{\emph{Harvard Journal of Law \& Technology}}
  \bibinfo{volume}{29}, \bibinfo{number}{2} (\bibinfo{year}{2016}),
  \bibinfo{pages}{353--400}.
\newblock
\urldef\tempurl%
\url{http://jolt.law.harvard.edu/articles/pdf/v29/29HarvJLTech353.pdf}
\showURL{%
\tempurl}


\bibitem[\protect\citeauthoryear{Stone, Brooks, Brynjolfsson, Calo, Etzioni,
  Hager, Hirschberg, Kalyanakrishnan, Kamar, Kraus, Leyton-Brown, Parkes,
  Press, Saxenian, Shah, Tambe, and Teller}{Stone et~al\mbox{.}}{2016}]%
        {stone_artificial_2016}
\bibfield{author}{\bibinfo{person}{Peter Stone}, \bibinfo{person}{Rodney
  Brooks}, \bibinfo{person}{Erik Brynjolfsson}, \bibinfo{person}{Ryan Calo},
  \bibinfo{person}{Oren Etzioni}, \bibinfo{person}{Greg Hager},
  \bibinfo{person}{Julia Hirschberg}, \bibinfo{person}{Shivaram
  Kalyanakrishnan}, \bibinfo{person}{Ece Kamar}, \bibinfo{person}{Sarit Kraus},
  \bibinfo{person}{Kevin Leyton-Brown}, \bibinfo{person}{David Parkes},
  \bibinfo{person}{William Press}, \bibinfo{person}{AnneLee Saxenian},
  \bibinfo{person}{Julie Shah}, \bibinfo{person}{Milind Tambe}, {and}
  \bibinfo{person}{Astro Teller}.} \bibinfo{year}{2016}\natexlab{}.
\newblock \bibinfo{booktitle}{\emph{Artificial {Intelligence} and {Life} in
  2030}}.
\newblock \bibinfo{type}{{T}echnical {R}eport}. \bibinfo{institution}{Stanford
  University}, \bibinfo{address}{Stanford, CA}.
\newblock
\urldef\tempurl%
\url{http://ai100.stanford.edu/2016-report}
\showURL{%
\tempurl}


\bibitem[\protect\citeauthoryear{Vinuesa, Azizpour, Leite, Balaam, Dignum,
  Domisch, Felländer, Langhans, Tegmark, and Nerini}{Vinuesa
  et~al\mbox{.}}{2019}]%
        {vinuesa_role_2019}
\bibfield{author}{\bibinfo{person}{Ricardo Vinuesa}, \bibinfo{person}{Hossein
  Azizpour}, \bibinfo{person}{Iolanda Leite}, \bibinfo{person}{Madeline
  Balaam}, \bibinfo{person}{Virginia Dignum}, \bibinfo{person}{Sami Domisch},
  \bibinfo{person}{Anna Felländer}, \bibinfo{person}{Simone Langhans},
  \bibinfo{person}{Max Tegmark}, {and} \bibinfo{person}{Francesco~Fuso
  Nerini}.} \bibinfo{year}{2019}\natexlab{}.
\newblock \bibinfo{title}{The role of artificial intelligence in achieving the
  Sustainable Development Goals}.
\newblock
\newblock
\showeprint[arxiv]{cs.CY/1905.00501}


\bibitem[\protect\citeauthoryear{Wallach and Marchant}{Wallach and
  Marchant}{2018}]%
        {wallach_agile_2018}
\bibfield{author}{\bibinfo{person}{Wendell Wallach} {and}
  \bibinfo{person}{Gary~E Marchant}.} \bibinfo{year}{2018}\natexlab{}.
\newblock \showarticletitle{An {Agile} {Ethical}/{Legal} {Model} for the
  {International} and {National} {Governance} of {AI} and {Robotics}}. In
  \bibinfo{booktitle}{\emph{Proceedings of the 2018 {AAAI} / {ACM} {Conference}
  on {Artificial} {Intelligence}, {Ethics} and {Society}}}.
  \bibinfo{publisher}{AAAI}, \bibinfo{address}{Palo Alto, CA},
  \bibinfo{pages}{7}.
\newblock
\urldef\tempurl%
\url{https://www.aies-conference.com/2018/contents/papers/main/AIES_2018_paper_77.pdf}
\showURL{%
\tempurl}


\bibitem[\protect\citeauthoryear{Weaver}{Weaver}{2014}]%
        {weaver_autonomous_2014}
\bibfield{author}{\bibinfo{person}{John~Frank Weaver}.}
  \bibinfo{year}{2014}\natexlab{}.
\newblock \showarticletitle{Autonomous {Weapons} and {International} {Law}:
  {We} {Need} {These} {Three} {International} {Treaties} to {Govern}
  “{Killer} {Robots}”}.
\newblock \bibinfo{journal}{\emph{Slate Magazine}} (\bibinfo{date}{Dec.}
  \bibinfo{year}{2014}).
\newblock
\urldef\tempurl%
\url{https://slate.com/technology/2014/12/autonomous-weapons-and-international-law-we-need-these-three-treaties-to-govern-killer-robots.html}
\showURL{%
\tempurl}


\bibitem[\protect\citeauthoryear{Zelli}{Zelli}{2011}]%
        {zelli_fragmentation_2011}
\bibfield{author}{\bibinfo{person}{Fariborz Zelli}.}
  \bibinfo{year}{2011}\natexlab{}.
\newblock \showarticletitle{The fragmentation of the global climate governance
  architecture}.
\newblock \bibinfo{journal}{\emph{Wiley Interdisciplinary Reviews: Climate
  Change}} \bibinfo{volume}{2}, \bibinfo{number}{2} (\bibinfo{year}{2011}),
  \bibinfo{pages}{255--270}.
\newblock
\showISSN{1757-7799}
\urldef\tempurl%
\url{https://doi.org/10.1002/wcc.104}
\showDOI{\tempurl}


\bibitem[\protect\citeauthoryear{Zelli and Van~Asselt}{Zelli and
  Van~Asselt}{2013}]%
        {zelli_introduction:_2013}
\bibfield{author}{\bibinfo{person}{Fariborz Zelli} {and} \bibinfo{person}{Harro
  Van~Asselt}.} \bibinfo{year}{2013}\natexlab{}.
\newblock \showarticletitle{Introduction: {The} {Institutional} {Fragmentation}
  of {Global} {Environmental} {Governance}: {Causes}, {Consequences}, and
  {Responses}}.
\newblock \bibinfo{journal}{\emph{Global Environmental Politics}}
  \bibinfo{volume}{13}, \bibinfo{number}{3} (\bibinfo{year}{2013}),
  \bibinfo{pages}{1--13}.
\newblock


\end{thebibliography}
\clearpage

\begin{appendix}
\section{Summary of Considerations}
\begin{table}[!htbp]
\renewcommand{\arraystretch}{1.5}
\centering
\begin{tabular}{|p{23mm}|p{25mm}|p{57mm}|p{57mm}|}\hline
{\bf Consideration} &
{\bf Implications for
\newline Centralisation} & {\bf Historical Example} & {\bf AI Policy Issue Example} \\ \hline
Political Power & Pro & {\it Shaping other regimes:} WTO has created a chilling effect, where the fear of conflicting with WTO norms and rules has led environmental treaties to self-censor to avoid addressing trade-related measures. & Empowered regime using foresight on AI systems development can address policy problems more quickly.
 \\ \cline{1-4}

Efficiency \newline \& Participation
 & Pro & {\it Decentralisation raises inefficiencies and barriers:} The proliferation of multilateral environmental agreements poses costs and barriers to participation in negotiation, implementation, and monitoring. & AI companies engage and share expertise, but if not checked by adversarial civil society, there is a greater concern of regulatory capture; increased costs undermine civil society participation.
 \\ \cline{1-4}

Slowness \newline \& Brittleness & Con & {\it Slowness:} Under the GATT, 1947 tariff  negotiations among 19 countries took 8 months. The Uruguay round, beginning in 1986, took 91 months for 125 parties to agree on reductions.
\newline \newline
{\it Regulatory capture:} WHO accused offor undue corporate influence in response to 2009 H1N1 pandemic.
\newline \newline
{\it Pathology of path-dependence:} Failed ILO reform attempts.

& Process of centralised regime can not keep pace with high speed of AI progress and deployment, may miss the window of opportunity.
\newline

Advanced AI issues (especially HLMI) may rapidly shift the risk landscape or problem portfolio of AI, beyond the narrow scope of an older institutional mandate

 \\ \cline{1-4}
 Breadth vs. Depth Dilemma & Con & {\it Watering down:} 2015 Paris Agreement suggest attempts to `get all parties on board' to centralized regime may result in significant watering down. & Attempts to effectively  govern the military uses of AI have been resisted by the most powerful states. 
\newline \newline
Attempted to create an IPAI have been resisted by the US and shifted to a smaller forum (the OECD).

\\ \cline{1-4}
Forum \newline Shopping
 & Depends on \newline design & {\it Power predicts outcomes:} \newline Intellectual property in trade shifted from UNCTAD to WIPO to WTO, with developed countries getting their way.
 \newline \newline
 {\it Accelerates progress:} NGOs and some states shifted discussions of anti-personnel mines ban away from CCW, ultimately resulting in the Ottawa Treaty. & Governance of military AI systems is fractured across CCW, multiple GGEs. This strategy may catalyze progress, but brings risks of fracture.
 \\ \cline{1-4}
 Policy \newline Coordination & Depends on \newline design & {\it Strong, but delayed convergence:} \newline Diverse regimes can coalesce into centralized regime, as seen with GATT and numerous trade treaties coalescing into the WTO, but doing so may take many decades. 

 & Numerous AI governance initiatives display loose coordination, but it is unclear if these initiatives can respond to policy developments in a timely manner.
 \\ 
\hline
\end{tabular}
\label{table2}
\end{table}
\end{appendix}

\end{document}